# The 1908 Tunguska Event and Bright Nights

Andrei Ol’khovatov

https://orcid.org/0000-0002-6043-9205
Independent researcher
(Retired physicist)
Russia, Moscow
email: olkhov@mail.ru

**Dedicated to the blessed memory of my grandmother ( Tuzlukova Anna Ivanovna ) and my mother ( Ol’khovatova Olga Leonidovna )**

**Abstract.** This paper is a continuation of a series of works, devoted to various aspects of the June 30, 1908 Tunguska event. In those days, various sky optical anomalies were observed over a large area. In the presented paper, the main focus is on bright nights. Bright nights are the glow (airglow) of the night sky, visible to the naked eye. In the author’s opinion, atmospheric gravitational waves played a large role in their appearance in 1908, as evidenced by the behavior of atmospheric pressure in late June and early July 1908. Some other geophysical peculiarities of that period, which could have affected the appearance of bright nights, and possibly other optical sky anomalies, are also considered.

## 1. Introduction

This paper is a continuation of a series of works in English, devoted to various aspects of the 1908 Tunguska event [Ol’khovatov, 2003; 2020a; 2020b; 2021; 2022; 2023a; 2023b].

More than a century has passed since the Tunguska event. A large number of hypotheses about its causes have already been put forward. However, so far none of them has received convincing evidence. In this paper the data concerning optical anomalies of a night sky around June 30, 1908 is considered with accent on bright nights. Bright nights are the glow (airglow) of the night sky, visible to the naked eye. .

While discussimg 1908-records from Russia, it should be remembered that according to the old Russian calendar the Tunguska event took place on June 17.

Please pay attention that so called the epicenter of the Tunguska forest-fall (the forest-fall is named “Kulikovskii”) is assigned to 60°53’ N, 101°54’ E. The author of this paper (i.e. A.O.) for brevity is named as “the Author”.

In the Author’s opinion, Igor Zotkin from KMET (the Committee on Meteorites of the USSR Academy of Sciences) and a group of Tomsk researchers led by Nikolai Vasil’ev (in some publications translated as “Vasilyev”) made the greatest contribution to the analysis of anomalies of the sky of that time. Vasil’ev knew several European languages and could read many publications on the subject in a library. Therefore, already in the late 1950s Tomsk researchers were very knowledgeable in this field. Here is a typical example. Here is a fragment from [Erokhovets, 1960] of a discussion in 1959 between KSE-1 (Kompleksnaya Samodeyatel’naya Ekspeditsiya – a group of researchers from Siberia – from Tomsk in general) and a group of tourists from Moscow (students in chemistry). Both groups were preparing to visit the Tunguska epicenter. The discussion was written down by a journalist Aleksandr Erokhovets, who was a member of KSE-1 (translated by A.O.):

“— Moreover, it is interesting, — said Gennady {Plekhanov – A.O.}, — that the glow was faintly noted on the eve of the fall. Here’s a new mystery.

— We haven’t heard about it, — the Muscovites were surprised.
— I have records with me, — said Lenya Shikalov.
— Give them here.
— Here is a description of a light phenomenon that occurred on June 17 in Timsky district, Kursk province, the Manturovo settlement.

> “... This phenomenon was noticed earlier, on June 15th and 16th, but ... to a lesser extent. On the 16th, however, quite a lot was observed effective lighting. The southern sky was especially good on the evening of June 16th, which took on a blue-azure hue and was pierced by golden streams of lightnings. It was surrounded by beautiful undulating, then pale golden, then dark gray clouds.”

— A strange glow, — the Muscovite girl said thoughtfully.
— What is important here: the sky shone from the south on the eve of the fall, the side is exactly where the body came from.
— Yes, it is difficult to explain this phenomenon only by the scattering of a meteorite dust released into the upper atmosphere, — Valery {Kuvshinnikov – A.O.} said. — In this case, the glow would have to be limited to the morning and evening dawns/glows, which have a reddish hue. This, for example, was observed when volcanic ash was released into the atmosphere when the Krakatoa volcano exploded in 1883... If these were meteoritic particles, then the glow would have to weaken gradually and continue at least since a year, because during this time the dust settles on the ground. But after the explosion of the Tunguska meteorite, the glow stopped after a few days. This can be explained by the end of the decay of short-lived radioactive isotopes, under the influence of which the air glowed. To that of course, meteor dust would glow differently and at a shorter distance..."

That was the case in 1959. Later many people around the world helped to collect the data. So in 1965 a fundamental book was published [Vasil'ev et al., 1965] (its responsible/managing editor was Igor Zotkin).

## 2. Bright nights

Now let's consider so called bright nights which took place near June 30, 1908. In general there were various optical sky peculiarities. So let's start with general situation on the sky of that time.

Krinov wrote [Krinov, 1966]:

"On the first night after the fall of the Tunguska meteorite, i.e. from 30 June to 1 July 1908, and with lesser intensity on a few successive nights, extraordinary optical phenomena were observed in the Earth's atmosphere."

The late evening of June 30, 1908 was remembered by eyewitnesses, first of all, because of the colorful twilight, which in some places lasted all night. So D.D. Rudnev in the village of Muratovo (the Oryol province) captured the twilight with his photocamera at 0:30 on July 1, and then manually colored the picture. The shutter speed was 10 minutes. This picture is shown in Fig.1 taken from [Nadein, 1909]. Since Muratovo is a very common name for a settlement, the Author can only say that it is most likely the village of Muratovo with coordinates about 53.2° N, 35.8° E.

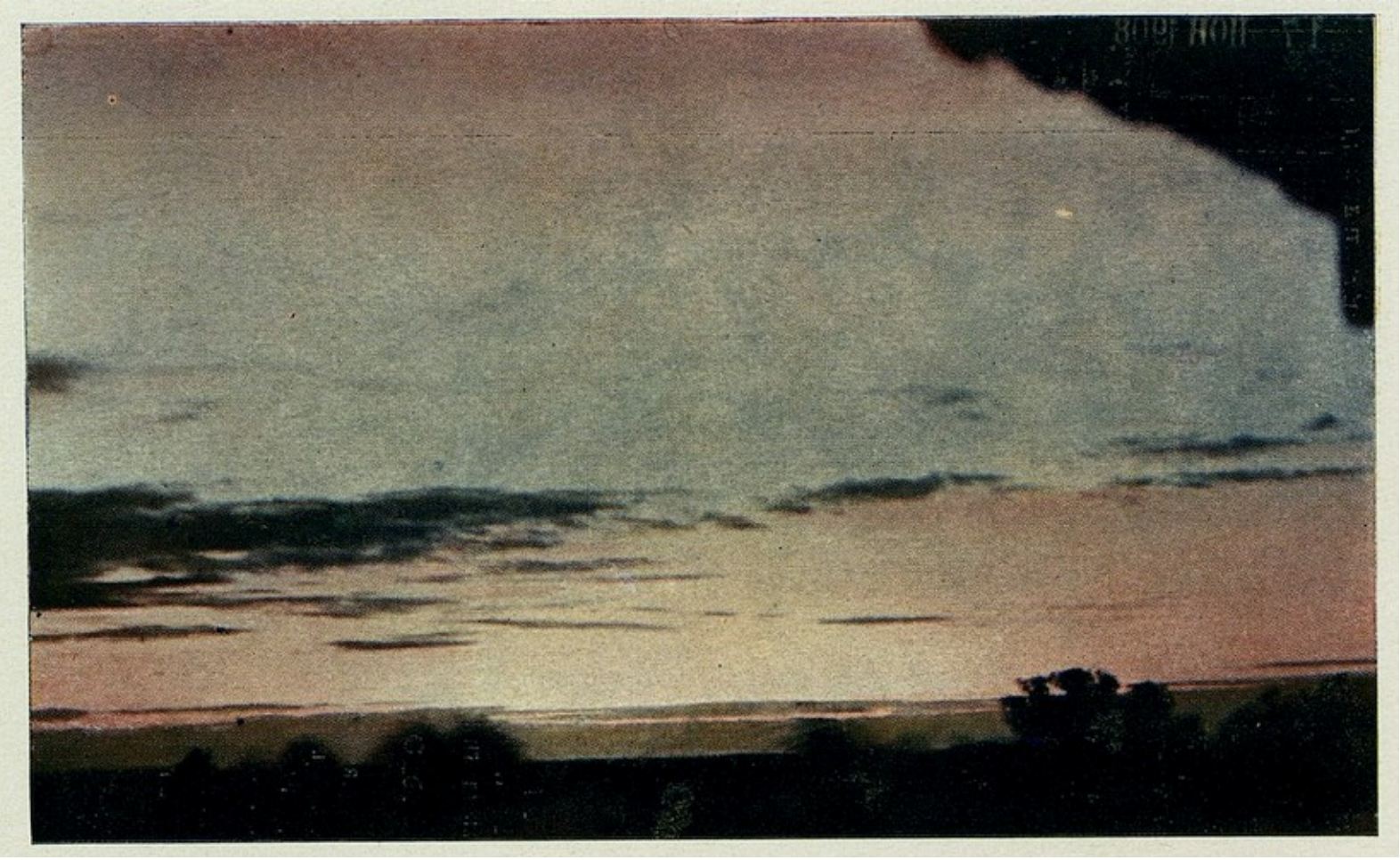

**Fig.1**

However the phenomena commenced before June 30. Here is from [Vasilyev, 1998]:

> “A detailed analysis of dynamics of June-July 1908 optical anomalies gives ground to a supposition that their first signs had appeared already several days before the fall of the meteorite: Suring (1908) points out that they began on 23 June, F. de Roy (1908) about 25 June and Denning (1908a, b) on 29 June. On that last day they were registered at 8 points in Germany, Holland, Britain, Sweden, Poland and Russia (Fig. 6). Nevertheless, the events reached their maximum (observed at more than 140 points) on the night of 30 June to 1 July.<…> Beginning with 1 July they vanished exponentially, but post-effects continued until the end of July 1908.”

Fig.6 is very close to the adapted (by A.O.) version from [Vasil’ev, 2004] on Fig.2 below. On Fig.2 along the vertical axis: total number of points where optical anomalies were registered — solid line (1), number of points where noctilucent clouds were observed—dashed line (2).

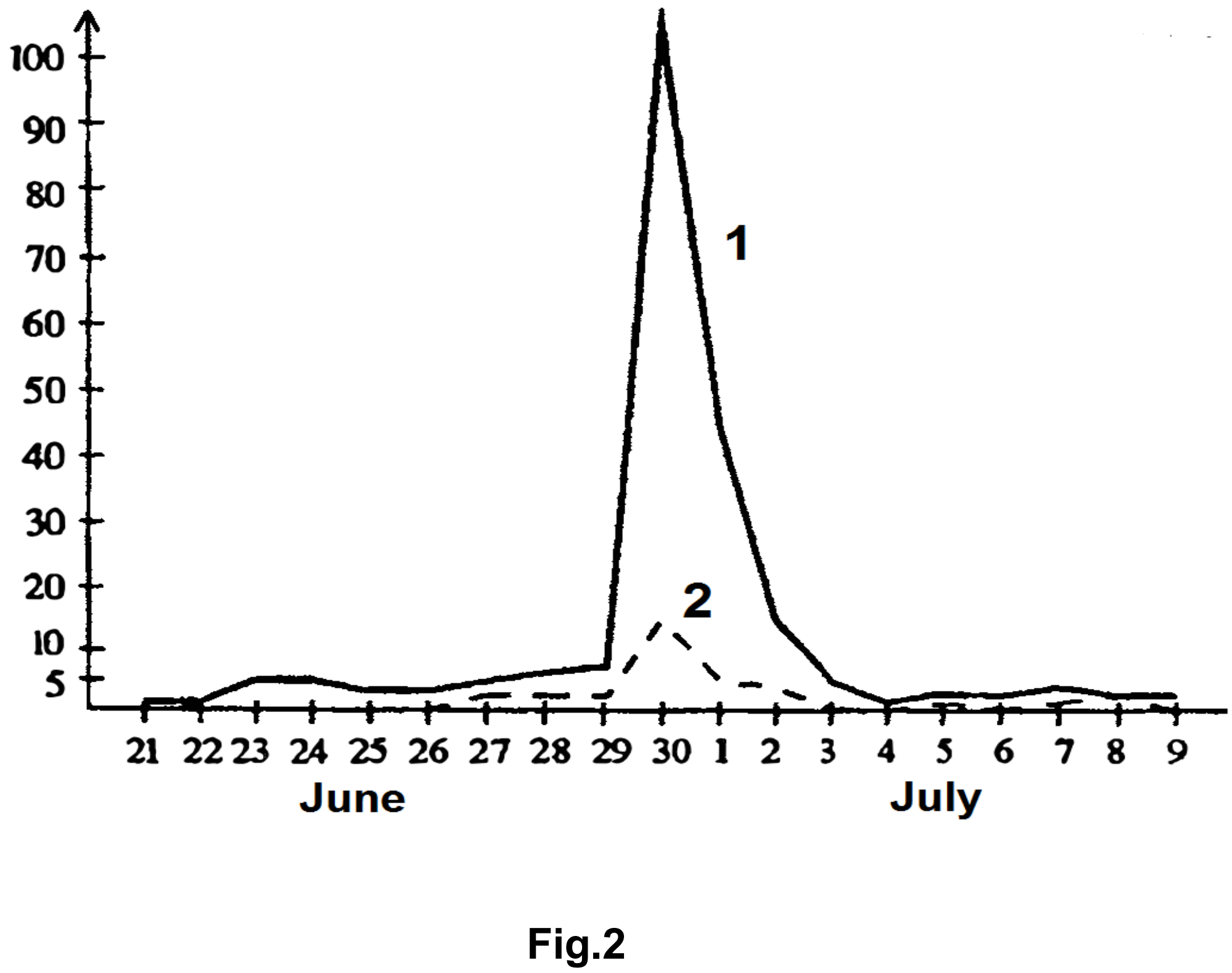

**Fig.2**

According to Vasil'ev the anomalies included abnormal sky-glows, unprecedented bright noctilucent clouds, disturbances in the normal run of Arago and Babinet neutral points and the appearance of unusually intense and prolonged solar halos. According to [Vasilyev, 1998] the area of anomalies was limited by the Yenisei River in the east, line Tashkent-Stavropol-Sevastopol-Bordeaux in the south and the Atlantic shore in the west. The northern border merged with the area of "white nights" usual at that season. It is possible to add that in 2008 new accounts were published (from an archive) that bright nights (moreover, they began before June 30) were also to the east of the Yenisei River – in the settlement of Sulomay area [Ol'khovatov, 2020a].

Vasil'ev wrote [Vasil'ev, 2004], translated by A.O.:

> "Let's add that a number of authors (who have described the anomalies of summer 1908 on the "hot trail") there was no doubt that the beginning of the anomalous period of the summer of 1908 refers to the twentieth of June.
>
> So, Felix de Roy wrote (1908): "Twilights, unusual as both in duration and distribution, they were observed throughout northern Europe, at least from the 45th parallel, in the last days of June and the first days of July 1908. These twilights appears to have appeared around June 25th. Having suddenly increased their intensity on the evening of June 30, they were still visible on July 1, weakening very quickly in the future."
>
> R. Süring came to a similar conclusion (Süring R., 1908): "The available reports prove that the extraordinary twilight phenomena are not limited to the two or three evenings mentioned so far in publications, but that they appeared as early as June 23 and with some interruptions, probably mainly due to cloudy conditions, were visible until the end of July" ."

Indeed, here is below on Fig.3 a fragment of the mentioned article by F. de Roy [de Roy, 1908].

## Conclusions générales.

Du matériel d'observations réuni jusqu'à présent se dégagent déjà, semble-t-il, quelques conclusions générales qu'on peut résuner comme suit :

1° Des crépuscules extraordinaires tant sous le rapport de la durée que de celui de l'étendue ont été observés dans tout le nord de l'Europe, tout au moins depuis le 45me parallèle, pendant les derniers jours de juin et les premiers jours de juillet 1908.

2° Ces crépuscules paraissent être apparus vers le 25 juin. Ils ont subitement augmenté d'intensité le soir du 30 juin, ont encore été remarqués le 1er juillet, et ont diminué très rapidement ensuite.

3° Sur le 50me parallèle environ, par ciel pur, un segment coloré s'étendait sur une longueur de 90° environ en azimuth et de 40° environ en hauteur. Il était régulier et se fondait avec le bleu du ciel par une zone de transition verdâtre. A l'intérieur de ce segment il s'en trouvait un second, plus ou moins rouge, et de 10 à 15° environ de hauteur.

4° Le point le plus coloré de ce segment se déplaçait pendant la nuit de l'E. vers l'W en suivant le cours apparent du soleil.

5° L'arc crépusculaire était assez vif pour illuminer tout le ciel et permettre, à minuit, de lire une montre. La Voie Lactée était invisible.

6° Les circonstances météorologiques locales, telles que ciel plus ou moins transparent, nuages, etc. influaient sur l'aspect de l'apparition, qui en était indépendante.

**Fig.3**

Also here is below on Fig.4 a fragment of the mentioned article by the prominent meteorologist Reinhard Süring [Süring, 1908].

Durch die hier vorliegenden Mitteilungen ist vor allem erwiesen, daß sich die ungewöhnlichen Dämmerungserscheinungen nicht auf zwei oder drei Abende beschränkt haben, von denen die bisherigen Veröffentlichungen allein sprechen, sondern daß sie schon seit dem 23. Juni aufgetreten und mit mannigfachen, hauptsächlich wohl durch die Bewölkungsverhältnisse bedingten Unterbrechungen bis Ende Juli sichtbar gewesen sind. Aus Schwarmitz a/Oder wird vom Abend des 23. Juni berichtet:

**Fig.4**

An outstanding meteorologist Wladimir Köppen wrote an article which title (translated as "Abnormal twilights on June 29 to July 1, 1908") speaks for itself [Köppen, 1908]. Here is on Fig.5 a fragment of the article.

**Abnorme Dämmerungen am 29. Juni bis 1. Juli 1908.** Daß die Lichterscheinung am Abend des 30. Juni keinem Nordlicht angehörte, wird in direkter Weise durch eine Beobachtung meines Kollegen Prof. Stück in Hamburg bewiesen. Der Genannte nahm am 30. Juni $11^{1}/_{2}{}^{p}$ in rechtweisend Nord eine starke rötliche Färbung des Himmels wahr, die anfangs den Eindruck eines hellen, durch Wolkenlücken sichtbaren, Nordlichts hervorrief. Eine sofortige spektroskopische Prüfung ergab ihm aber unzweideutig, daß die Erscheinung kein Polarlicht, sondern eine ungewöhnliche Dämmerungserscheinung war.

**Fig.5**

Arthur Stentzel (who systematically carried out observations of twilights in the vicinity of Hamburg in general) noted that the appearance of abnormally bright twilights should be attributed to June 22, 1908, and that the unusual color tones of the twilights have been taking place since April. He ended his article [Stentzel, 1909] with words that we do not need to use cosmic factors to explain it (at those times cosmic dust was usually considered). Here is a fragment of the article [Stentzel, 1909] on Fig.6.

446 Meteorologische Zeitschrift. Oktober 1909.

Nächte beobachtete ich vom 30. Juni bis zum 12. Juli 1908 täglich, in den folgenden Tagen hinderten Wolken die Beobachtung der Erscheinung, sie kehrte wieder am 19., 21. und 22. Juli, blieb dann aber endgültig aus. Die mehrfach behauptete Plötzlichkeit der Erscheinung war in Wirklichkeit nicht in dem Maße vorhanden, das Phänomen bereitete sich vielmehr, wie aus der Tabelle leicht ersichtlich ist, bereits vom 22. Juni an vor.

| | 7 | 6 | 5 | 4 | 3 | 2 | 1 |
|---|---|---|---|---|---|---|---|

**Fig.6**

The above represents only a small part of the data collected by the Vasil'ev's group.

Thus, the researchers "in hot pursuit" did not consider that the phenomena before June 30 were fundamentally different from subsequent phenomena, but were the initial stage of the phenomena's development. The development of these phenomena gradually increased, culminating on the night of June 30 to July 1 and fading in the following days. The difference in the definition of the start time for different authors is probably due to the gradual development of anomalies, when in different places and by different observers (especially taking into account clouds, etc.) the start could be fixed at different times. And if it was weakly manifested, that it might not be given much importance (at least to report it publicly).

According to [Vasil'ev et al., 1965] the optical anomalies of the summer of 1908 are a complicated complex, the main features of which are: 1) increase the night sky glow; 2) the appearance of unusually bright dawns/twilights; 3) powerful development of the noctilucent clouds; 4) changing the polarization properties of the daytime sky. Regarding these anomalies Vasil'ev with co-authors wrote [Vasil'ev et al., 1965] translated by A.O.:

> "We come to the conclusion that all the anomalies of the summer 1908 mentioned at the beginning of the paragraph are a single whole, a single complex, suggesting the presence of a single cause."

It was known (an empirical rule) that noctilucent clouds tend to occur during upsurge of atmospheric pressure [Vasil'ev et al., 1965]. The authors of [Vasil'ev et al., 1965] tried to test this on events around June 30, 1908, and got mixed results. In some areas the noctilucent clouds occurred without accompanying air-pressure upsurges.

In 1990s the Author looked at the data of meteorological stations in Russia of 1908 [Rykatchew, 1911] to discover possible connection between noctilucent clouds and atmospheric pressure in June-July of 1908. He discovered that the tendency took place indeed (only tendency, not 100% connection). Also the upsurges of air-pressure were repeated several times resembling waves.

The author also tried to investigate the data of meteorological stations in Russia in those times for possible connection between bright nights and atmospheric pressure in June-July of 1908. One of the problems was that it was often unclear from the descriptions of eyewitnesses whether there was an increase in the own luminosity of the clear sky (i.e. not caused by twilight, clouds, etc.). In a couple of cases (for which the Author has discovered weather data), such the luminosity increase was strongly suspected from the descriptions. But the number of cases was too small for any reliable conclusion.

So the Author wrote in [Ol'khovatov and Rodionov, 1999] (the book was written as a discussion between two its authors) regarding the air-pressure upsurges connected with noctilucent clouds (translated by A.O.):

> "Presumably, these were the so-called atmospheric Rossby waves, the source of which could be intense atmospheric cyclonic movements during that time period."

Since 1990s some progress was made. In 2017 an important article appeared [Shepherd and Cho, 2017]. The authors researched the longitudinal variations of $O(^1S)$ airglow emission rate obtained on a satellite. Their results are very interesting [Shepherd and Cho, 2017]:

> "The pattern generally appears as a combination of zonal waves 1, 2, 3, and 4 whose phases propagate at different rates. Sudden localized enhancements of 2 to 4 days duration are sometimes evident, reaching vertically integrated emission rates of 400 R, a factor of 10 higher than minimum values for the same day. These are found to occur when the four wave components come into the same phase at one longitude. It is shown that these highly localized longitudinal maxima are consistent with the historical phenomena known as "bright nights"…"

So their result states that the reason for bright nights is superposition of Rossby waves. This fact is a reason to look again at bright nights of 1908. Especially as since 1990s much weather data was placed in internet.

Let's start with bright nights reported from the Manturovo settlement (~51.5° N, ~37.1° E) - see early in this paper. There was a meteorological station (meteostation) in the city of Kursk about 70 km away. On Fig.7 along the vertical axis: air-pressure in mm Hg from [Rykatchew, 1911].

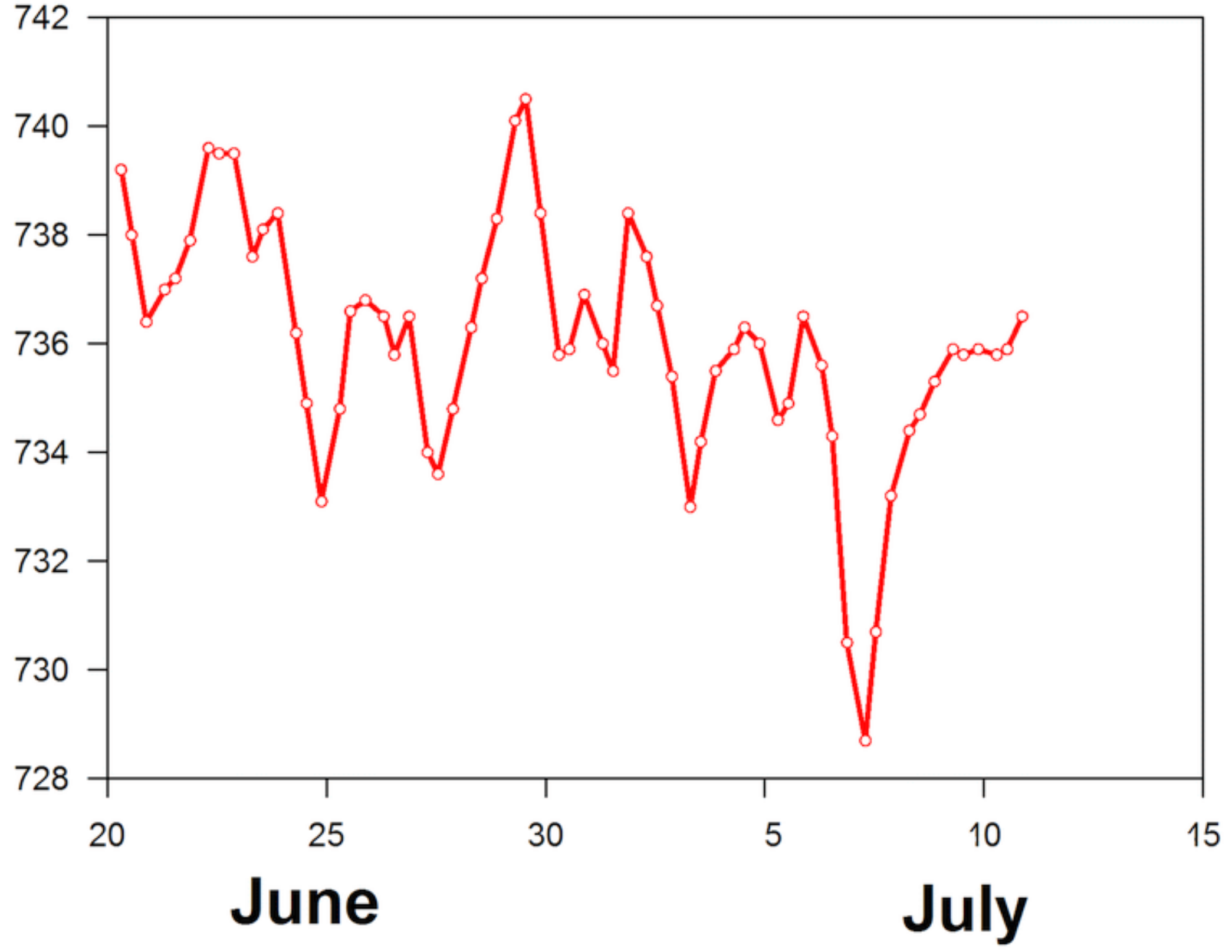

**Fig.7**

An upsurge on the days of bright nights is clear visible. Also the words about lightnings add one more aspect. Indeed the data from the Kursk meteostation shows distant thunderstorms on those days. A thunderstorm can generate atmospheric gravity waves. Gravity wave energy and momentum can be transported from thunderstorms into the upper atmosphere, and to result in bright nights [Smith et al., 2020].

Unlike Kursk, a meteostation in Tashkent (41°20' N, 69°18' E, altitude 478.3 m) marked the fair weather in late June and early July of 1908. In the previous days, light clouds were noticed only on June 26, and the next time a slight cloud cover appeared only on July 6. V.G. Fesenkov recalled [Fesenkov, 1968] translated by A.O.:

> "I was at the Tashkent Observatory at that time and, together with the observatory's astrophysicist, I. I. Sikora {Josif Josifovič Sikora – A.O.}, I had to photograph the sky using a large astrograph. On the evening of June 30, we stood for a long time at the astrograph tower, waiting for darkness, but the sky remained evenly leaden-pale. We still did not see the stars. The night did not come. If this phenomenon was caused by the scattering of sunlight in a high atmosphere, then for the latitude of Tashkent, the scattering particles would have to stay at an altitude of about 800 km! It was something unusual and incomprehensible. On the following day, the brightness of the night sky decreased by several tens of times already and then quickly returned to normal.”

Here is on Fig.8 the air-pressure data from the Tashkent meteostation [Rykatchew, 1911]. On Fig.8 along the vertical axis: air-pressure in mm Hg from [Rykatchew, 1911].

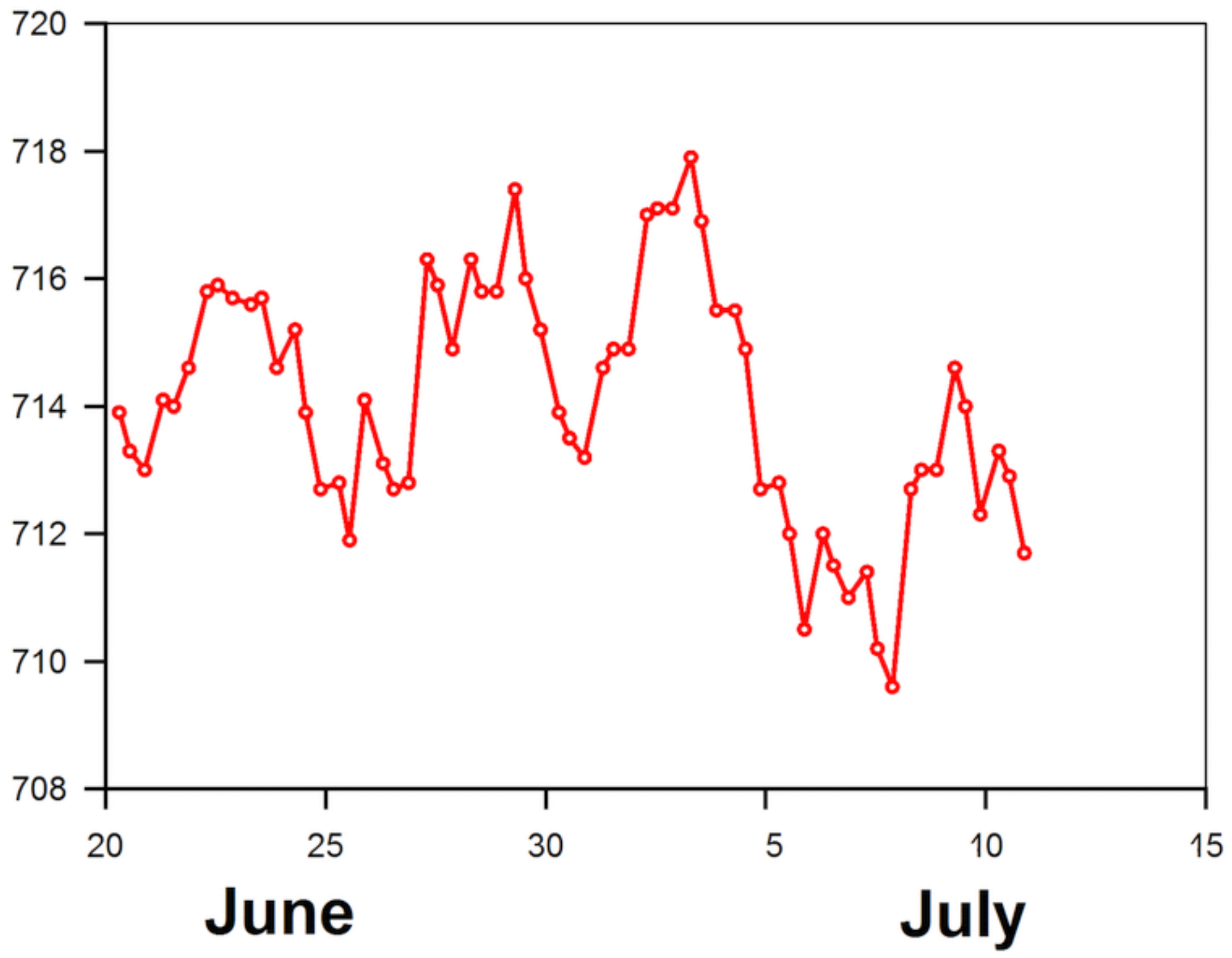

**Fig.8**

Interestingly that in Tashkent bright night took place during a final phase of an air-preesure upsurge – i.e. during fast decrease of air-pressure.

Here is a fragment from a note by a prominent British meteor observer William Frederick Denning, who was in the city of Bristol area [Denning, 1908]:

“Here the display ranged over four nights, for on June 29 the sky was very light, and stars and Milky Way extremely faint, but clouds were very prevalent. On July 2 some attractive, coloured-cloud scenery was presented in the north-west and north, but the sky had not the bright, weird aspect it wore on preceding nights, and after midnight I saw nothing unusual.”

The Author discovered scanned weather info of meteostation in the town of Bath (about 15 km from Bristol) on web-site of the UK Meteorological Office ( https://www.metoffice.gov.uk ). On Fig.9 along the vertical axis: air-pressure (inches, reduced to M.S.L. and 0 °C).

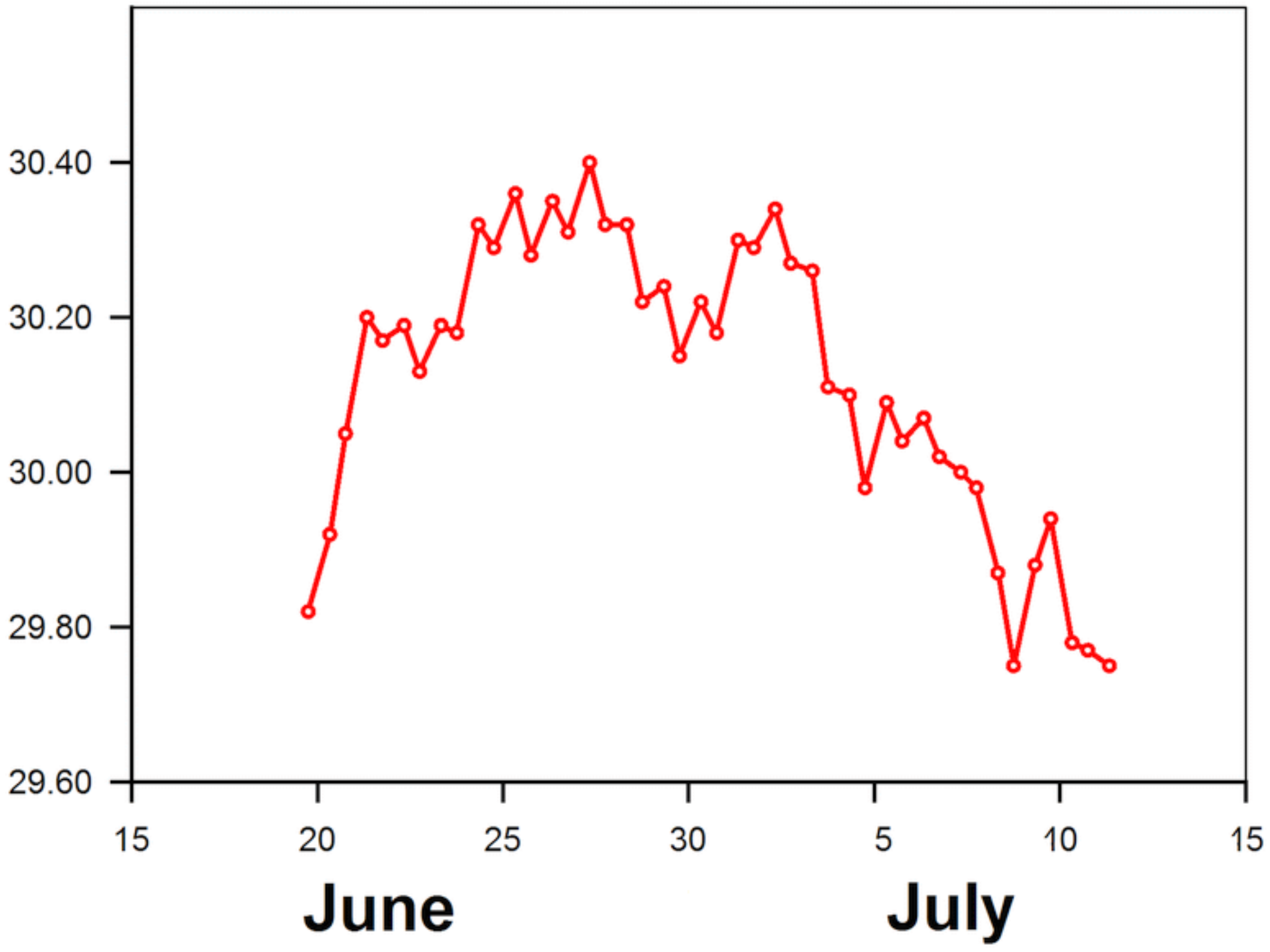

**Fig.9**

One more place with bright night in 1908 probably was the German town Heidelberg, where a prominent astronomer Max Wolf saw that even far beyond the zenith, the sky was still quite brightly lit, even as far as the southern horizon [Wolf, 1908].

The Author discovered scanned weather info of meteostation in the city of Frankfurt (about 80 km to the north from Heidelberg) on web-site of the UK Meteorological Office ( https://www.metoffice.gov.uk ). On Fig.10 along the vertical axis: air-pressure (inches, reduced to M.S.L. and 0 °C).

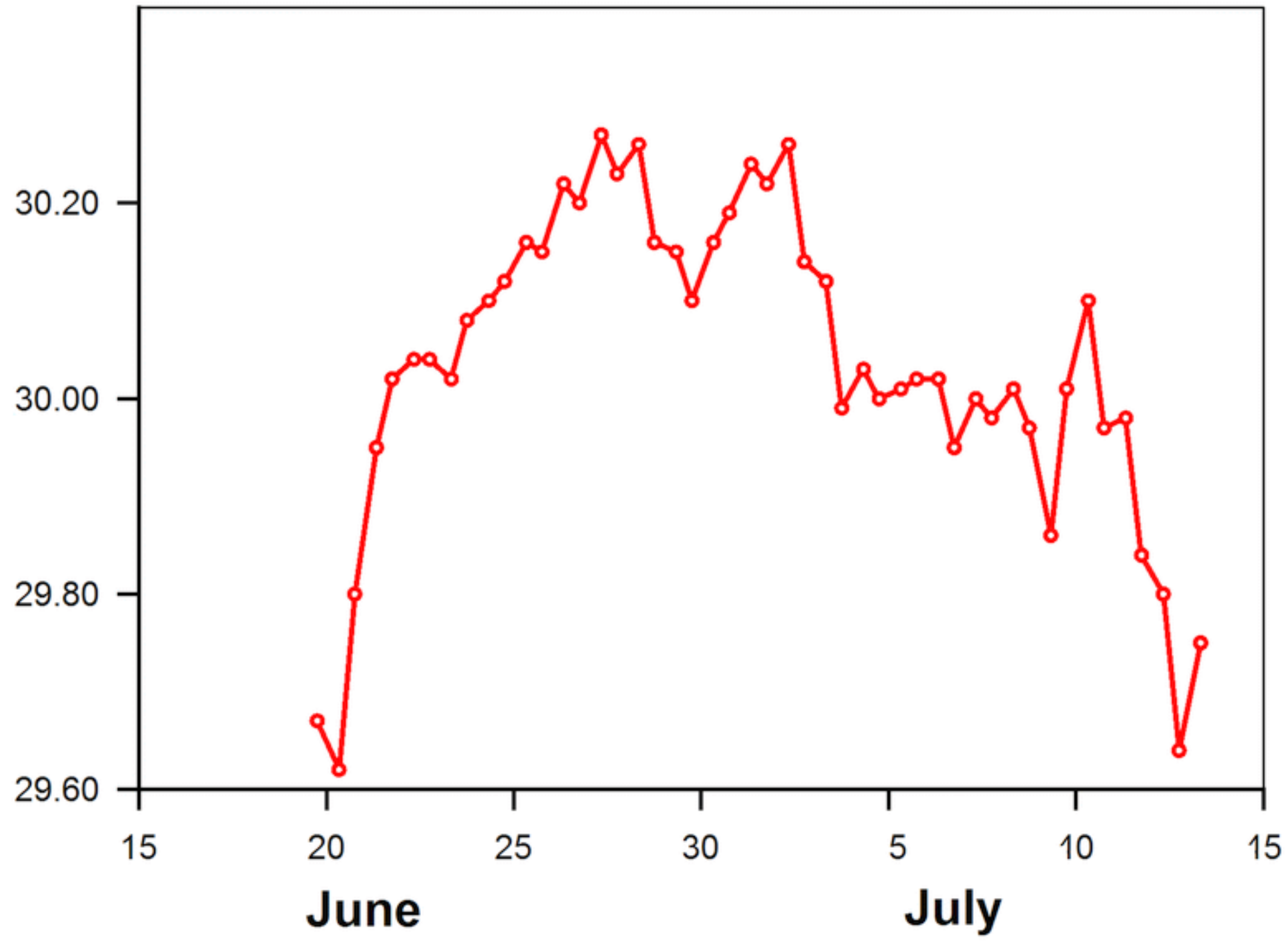

**Fig.10**

In the city of Cracow a student Joh. Krassowski also saw the greenish glow from an astronomical observatory on the late evening of June 30 [Rudzki, 1908]. Here is the fragment translated from German by Peter C. Slansky:

> "At 11h p.m. an isolated arc-shaped streak could be seen about 5° above the most westerly side of the sector, the concavity of which was turned towards the sector. The boundaries of the sector were quite sharp, while the streak consisted of green, diffuse light.
>
> At about 11h p.m. $\alpha$ Aurigae (Capella) was located at the boundary of the sector. The star shone as a 3rd magnitude star. A quarter of an hour later the sky became very bright. On Earth, distant objects and bright parts of the Milky Way in the region of Cassiopeia could be seen, and I could not distinguish Cygnus from the celestial background. The greenish brightness extended to above the zenith, bright stars in the Lyra star cluster were as if dazzled.

At 12 p.m. the outlines of the sector became less sharp. In the whole N-sky the green light spread in broad radial stripes. It dazzled the light of the stars. I noticed, for example, that α and ß Ursae majoris looked like 3rd magnitude stars."

Unfortunately it is not clear how far the greenish brightness did extend.

The Author discovered scanned weather info of meteostation in the city of Cracow on web-site of NOAA ( https://www.noaa.gov ). On Fig.11 along the vertical axis: air-pressure in millimeters.

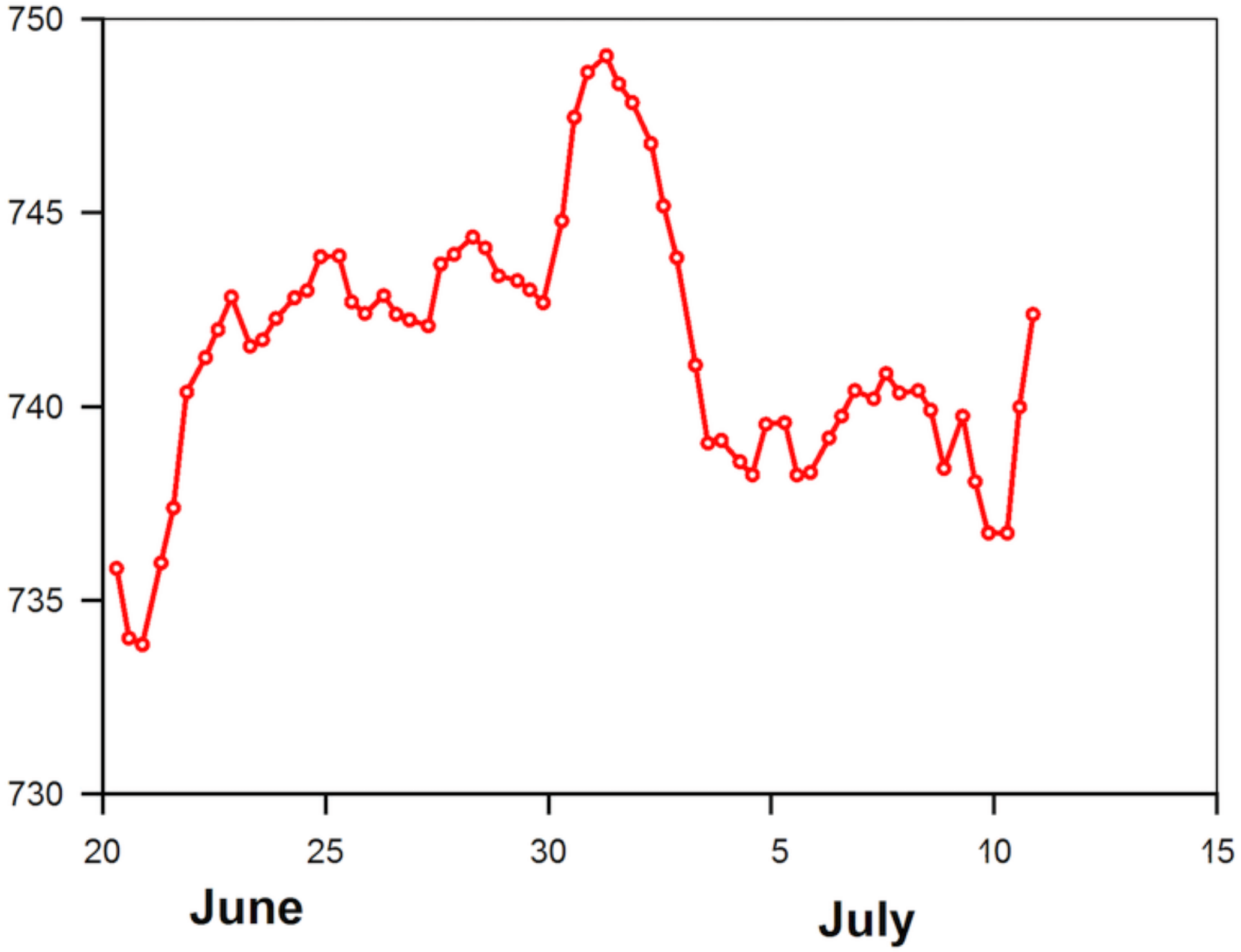

**Fig.11**

The graph shows a rapid increase in air-pressure since the morning of June 30.

At the end of December 1916, remarkable bright nights again took place in a large part of Europe [Mora, 1917]. In some places, the bright night from December 23 to December 24 attracted special attention, in other places from December 24 to December 25. But this happened during the First World War, so these phenomena have remained poorly researched. Let's look at the air-pressure data during the bright nights of 1916. The Author discovered scanned weather info of meteostation in the city of Paris on web-site of the UK Meteorological Office ( https://www.metoffice.gov.uk ). Unfortunately data for the evening of Jan. 6 and morning Jan.7 (1917) was absent. On Fig.12 along the vertical axis: air-pressure is given in milliBars at MSL.

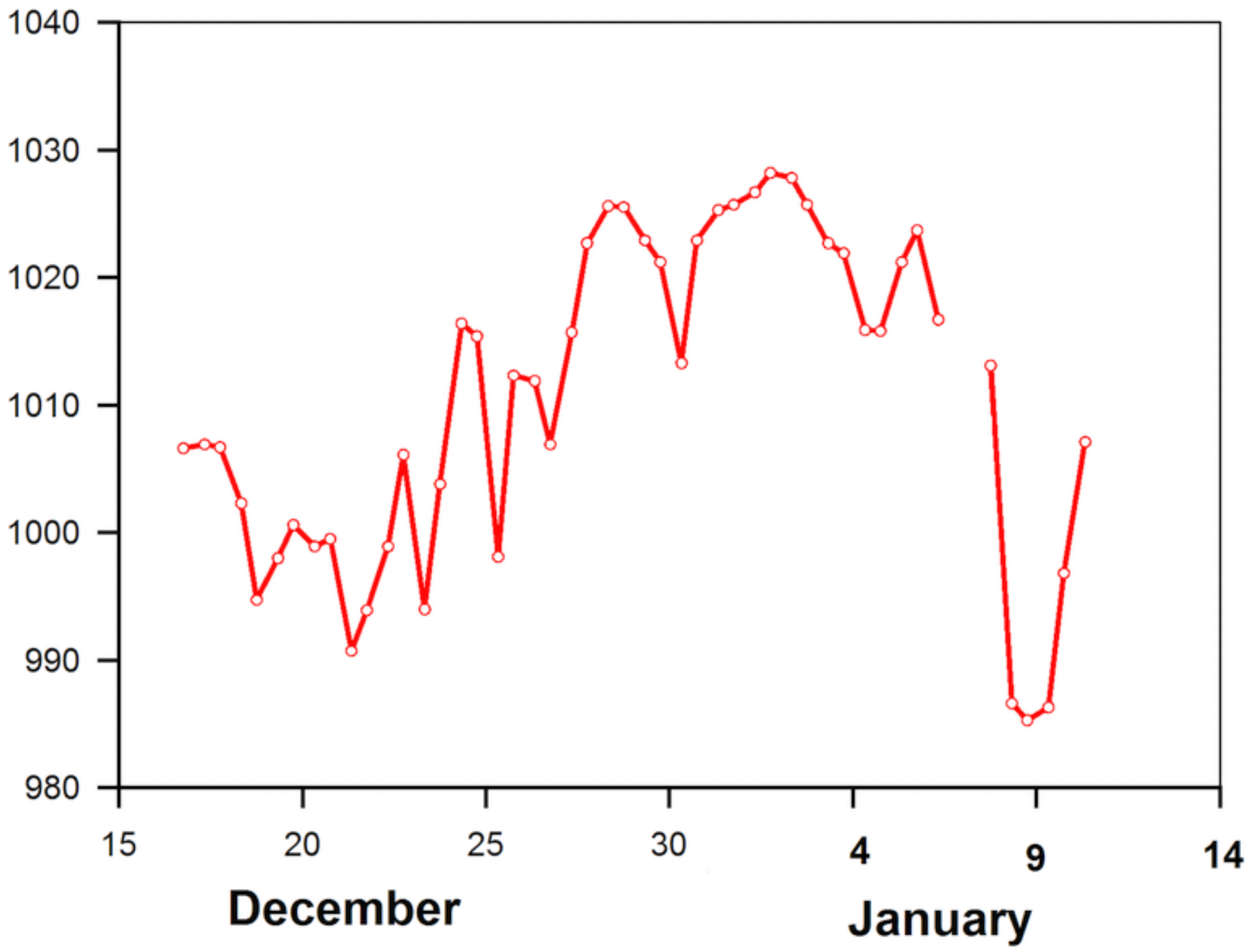

**Fig.12**

Bright nights in 1908 and 1916 were observed over a large region. Let's now consider another case, described in [Strutt, 1931]. Robert John Strutt (the 4th Lord Rayleigh) observed the bright night from Terling Place (51.8021°N, 0.57097°E ) on the night between November 8 and 9, 1929. He saw the glow with the naked eye [Strutt, 1931]:

> "...and on going out into the open the exceptional brightness was seen all over the visible hemisphere, no direction being obviously favoured."

The Author discovered scanned weather info of the Croydon meteostation (about 66 km from the place of observation) on web-site of the UK Meteorological Office ( https://www.metoffice.gov.uk ). Unfortunately, the scanned pressure reading at 01 GMT on November 5 is not very clear. It is possible that the value is 1028.2. On Fig.13 along the vertical axis: air-pressure is given in milliBars at MSL.

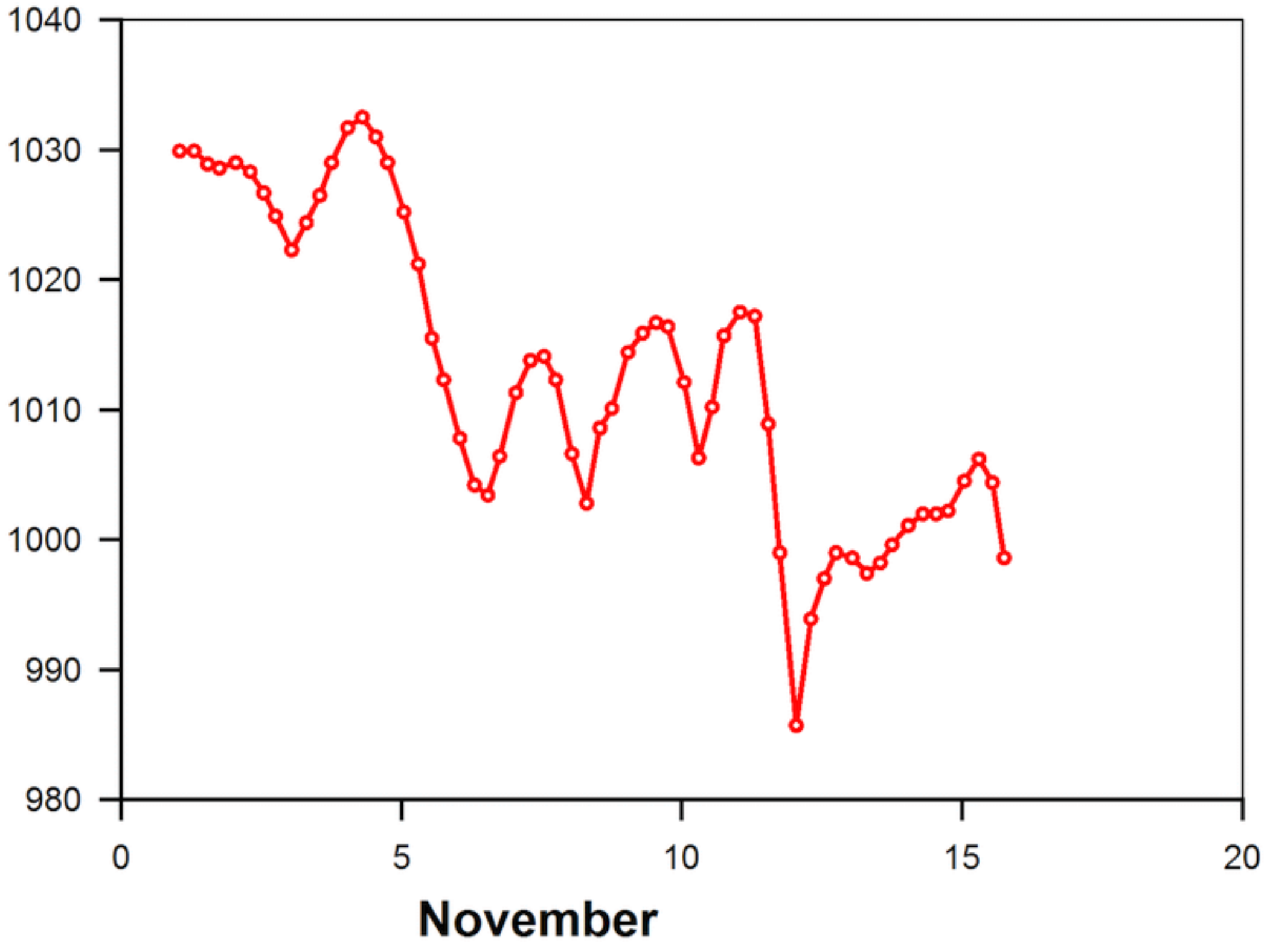

**Fig.13**

## 3. Discussion

As follows from the presented data, optical anomalies of the sky (in particular, bright nights) occurred in those days against the background of significant changes in air-pressure (this was especially pronounced in 1908). Indeed, these pressure changes hint at the energy and momentum flows into the upper atmosphere that existed in those days.

The question may arise: if the weather forecast predicts sharp changes in atmospheric pressure, can we expect bright nights? For now, the Author can answer this as follows. So far, we can only talk about an increase in the probability of more or less significant increased airglow. In addition, it is far from certain that this gain can be so significant that the airglow will become visible to the naked eye.

Indeed, air-pressure fluctuations occur quite often, and bright nights, especially at the 1908 level, are incomparably less frequent. Therefore, it can probably be assumed that these fluctuations and the associated flows of energy and momentum into the upper atmosphere play the role of a favorable (contributing) factor in the case of a "very" bright night at least. So for a "very" bright night, especially such as in 1908 (when other sky anomalies were also reported), some additional reasons are needed.

The Author in [Ol'khovatov, 1997] suggested several possible factors that could lead to the increased airglow. Here are some of them.

According to [Kozhenkova et al., 1963], translated by A.O.:

> "It is anomalous {in the first five days of July – A.O.} that the cyclones moving across ETC {the European part of the state/country –A.O.} and Western Siberia were deeper than the cyclones in the first region {North America, Greenland, the western Atlantic and the northern Pacific – A.O.}.
>
> <...>
>
> In comparison with the long-term average, in 1908 there was a sharp increase in pressure gradients over all regions of the Soviet Union. This is explained by the fact that the cyclones moving through this region were deep. This also led to negative pressure deviations from the norm; the deviation values ranged from 6-13 mm."

It is likely that such peculiarities of cyclonic activity led to the above-stated air-pressure upsurges/fluctuations and to the airglow in a certain region (for example, there were no reports of the airglow from North America).

According to [Yakovlev, 1990] translated by A.O.:

> "According to Russian meteorological stations, on June 29 and 30, 1908, there was an increased amount of precipitation with a maximum on June 30 and an increased number of stations with thunderstorms compared to neighboring dates. The total hours of sunshine after June 29th is falling."

In July 1908, there was also a slight decrease in the transparency of the Earth's atmosphere, which at one time was attributed to the dispersed substance of the alleged Tunguska meteorite. So in 1949, V.G. Fesenkov, based on the data [Abbot et al., 1913], estimated the order of mass of the alleged meteorite at least several million tons [Fesenkov, 1949]. The data [Abbot et al., 1913] was investigated again in [Turco et al., 1982], where it was written:

> "We deduce that nearly 1 million tons of pulverized dust may have been deposited in the mesosphere and stratosphere by the Tunguska fall, which agrees with previous estimates of the meteor mass influx."

However the Fesenkov's estimation seems to be a bit larger.

Anyway since the end of 1980s a new series of articles appears which gives another interpretation of the data published in [Abbot et al., 1913]. For example, an analysis of the spectral features of atmospheric attenuation conducted in [Nikol'skii and Shul'ts, 1990] revealed that the opacity was periodic in nature and began as early as May 1908. According to the authors, the probable cause was a giant meteorite or a group of them.

Based on the values of atmospheric transmission for different wavelengths, these researchers concluded that the alleged "Tunguska space body" did not introduce any noticeable amount of dust into the atmosphere and, in their opinion, consisted mainly of water-ammonia-methane ice weighing about 1000 Mt.

In [Nikol'skii and Shul'ts, 1990] it was suggested that since June 19, 1908, the moisture content of the atmosphere (especially in the upper layers) began to increase. According to the authors, the process of increasing humidity in the atmosphere continued until about August 8, 1908, bringing integral humidity to a level corresponding to 2.6 cm of precipitated water. The same authors reported that this phenomenon was no longer found in the data for the other seasons they studied.

In the Author's opinion the facts presented above indicate that very large-scale geophysical processes played a role in the appearance of bright nights in 1908. Some other large-scale (perhaps even global) geophysical peculiarities of that time period are discussed in [Ol'khovatov, 2003].

In [Ol'khovatov, 1997] the Author also considered some possible extraterrestrial factors. Let's look at some of them in more detail.

On June 28, 1908 an annular solar eclipse occurred. Thus, during the days of maximum bright nights, there should have been an increased tide in the upper atmosphere generated by the combined action of the Moon and the Sun.

By the way, processes in the Sun could have made a certain contribution. In [Levander and Maunder, 1908] there is an interesting phrase at the meeting of the British Astronomical association on July 1, 1908:

> “Prof. Fowler noted on the 30th that there was a very fine and bright prominence at the limb, and he mentioned that he thought there would be a display of aurora in the evening, and there had been one on the previous (Tuesday) night.”

On July 1 even more remarkable prominence was noticed. In [Daunt, 1909] which was a summary of prominences in 1908, the July 1 prominence was the only one in section “Dates of special prominence outbursts”. It was marked as “Not a large, but a bright and very active prominence”. The author in [Ol’khovatov, 1997] even suspected solar proton event, but in 1908 existence of protons was unknown. Brauner [Brauner, 1908] wrote from Prague:

> "It is reported that magnetic disturbances were experienced on the telegraphic lines, but I saw no trace of the characteristic auroral bands or columns."

Interestingly that there were not strong geomagnetic disturbances on those days [Vasil’ev et al., 1965], but on the other hand in Antarctic near the South magnetic pole an exceptional aurora was seen about seven hours prior to the Tunguska explosion [Steel and Ferguson, 1993]. Such controversial (at glance, at least) data hints that something unusual took place on our planet.

## 4. Conclusion

In the Author’s opinion, there are quite a lot of arguments in favor of the fact that the bright nights of 1908 were caused by the geophysical peculiarities of that time period.

In order to study such large-scale bright nights as of 1908, it is desirable to combine the efforts of meteorologists, specialists in atmospheric optics, upper atmosphere, solar physics, and possibly other specialties.

In recent decades, reports of bright nights have been very rare, one of the reasons for which is probably "light pollution" by lighting. Therefore, such reports are very valuable, especially for further study of the nature of this remarkable phenomenon.

**ACKNOWLEDGEMENTS**

The author wants to thank the many people who helped him to work on this paper, and special gratitude to his mother - Ol’khovatova Olga Leonidovna (unfortunately she didn’t live long enough to see this paper published...), without her moral and other diverse support this paper would hardly have been written.

## References

Abbot, C. , Fowle, F. , Aldrich, L. (1913). Results of measurements of the intensity of solar radiation.// Annals of the Astrophysical Observatory of the Smithsonian Institution, vol. 3, pp.73-113.

Brauner, Bohuslav (1908). The Recent Nocturnal Glows. // Nature, v. 78, N 2019, p. 221.

Daunt, R. (1909). Solar Prominences. Summary for 1908. // Journ. Brit. Astron. Assoc., v.19, N4, pp.162-164.

Denning, W.F. (1908). The Sky Glows. // Nature, vol.78, N 2020, p.247.

de Roy, Felix (1908). Les illuminations crepusculaires des 30 juin et 1er juillet 1908 . // Gazette Astronomique, vol. 1, N 8, pp.61-64 (in French).

Fesenkov, V. G. (1949). Pomutnenie atmosfery, proizvedennoe padeniem Tungusskogo meteorita 30 iyunya 1908 g. // Meteoritika, v.6, pp. 8-12 (in Russian).

Fesenkov, V. G. (1968). Tungusskoe yavlenie 1908 goda. // Zemlya I Vselennaya, N 3, pp. 4 – 10 (in Russian).

Erokhovets, A. (1960). Meteorit ili zvezdnyi korabl’? // Sibirskie ogni, N10, p.114 (in Russian).

Köppen, W. (1908). Abnorme Dämmerungen am 29. Juni bis 1. Juli 1908. // Meteorologische Zeitschrift,v.25, pp. 571 - 572 (in German).

Kozhenkova, Z.P., Brok, V.A., Fedyushina L.P., et al. (1963). Sinoptiko-meteorologicheskie usloviya leta 1908 g. // Problema Tungusskogo meteorita. Tomsk, Izdatel'stvo Tomskogo universiteta, pp.179 - 186 (in Russian).

Krinov, E.L. (1966). Giant meteorites. Pergamon Press, Oxford, London, New York, 397 p.

Levander, F. W. and Maunder, E. W. (1908). Report of the meeting of the Association Held on July 1st 1908, at Sion College, Victoria embankment, E.C. // Journ. Brit. Astron. Assoc., v.18, N9, pp. 349-355.

Mora, Enzo (1917). La nuit claire du 23-24 décembre 1916. // L'Astronomie, v. 31, pp. 309 - 310 (in French).

Nadein, I. (1909). Svetyaschiesya oblaka. // Zapiski po gidrografii, t.31, pp.381 -387 (in Russian).

Nikol'skii, G.A., Shul'ts, E. O. (1990). O svyazi povtoryayuschikhsya pomutnenii atmosfery v 1908 g. s vtorzheniem krupnykh kosmicheskikh tel. // Meteoritika, v. 49, pp. 202 – 218. (in Russian).

Ol'khovatov, Andrei (1997). Mif o Tungusskom meteorite. Tungusskii fenomen 1908 goda – zemnoe yavlenie. Assots. "Ekologiya Nepoznannogo", Moskva, 128 p. (in Russian).

Ol'khovatov, A.Yu. (2003). Geophysical Circumstances Of The 1908 Tunguska Event In Siberia, Russia. // Earth, Moon, and Planets 93, 163–173. https://doi.org/10.1023/B:MOON.0000047474.85788.01

Ol'khovatov, Andrei (2020a). New data on accounts of the 1908 Tunguska event.// Terra Nova,v.32, N3, p.234. https://doi.org/10.1111/ter.12453

Ol'khovatov, Andrei (2020b). Some comments on events associated with falling terrestrial rocks and iron from the sky. // https://arxiv.org/abs/2012.00686

https://doi.org/10.48550/arXiv.2012.00686

Ol'khovatov, Andrei (2021) - The 1908 Tunguska event and forestfalls. // eprint arXiv:2110.15193 ,  https://doi.org/10.48550/arXiv.2110.15193

Ol'khovatov, Andrei (2022) - The 1908 Tunguska Event And The 2013 Chelyabinsk Meteoritic Event: Comparison Of Reported Seismic Phenomena. // eprint arXiv:2206.13930 , https://doi.org/10.48550/arXiv.2206.13930

Ol'khovatov, Andrei (2023a) - The 1908 Tunguska event: analysis of eyewitness accounts of luminous phenomena collected in 1908. // arXiv:2310.14917

https://doi.org/10.48550/arXiv.2310.14917

Ol'khovatov, Andrei (2023b) - The Evenki accounts of the 1908 Tunguska event collected in 1920s – 1930s. // arXiv:2402.10900 ,
https://doi.org/10.48550/arXiv.2402.10900

Ol'khovatov, A.Yu., Rodionov, B. U. (1999). Tungusskoe siyanie. M., Laboratoriya Bazovykh Znanii, 240 p. (in Russian). ISBN 5-93208-027-2

Rudzki, M.P. (1908). Nordschein am 30. Juni in Krakau. // Meteorologische Zeitschrift, v.25, p. 313 (in German).

Rykatchew, M. (1911). Annales de l'observatoire physique central Nicolas, Annee 1908, Peterburg. (in French).

Shepherd, G. G. , Cho, Y. -M. (2017).WINDII airglow observations of wave superposition and the possible association with historical " bright nights " //
Geophysical Research Letters, v.44, pp.7036-7043,
https://doi.org/10.1002/2017GL074014

Smith, Steven M. , Setvák, Martin , Beletsky, Yuri et al. (2020). Mesospheric Gravity Wave Momentum Flux Associated With a Large Thunderstorm Complex. // Journal of Geophysical Research: Atmospheres, Volume 125, Issue 21, article id. e33381

https://doi.org/10.1029/2020JD033381

Steel, D. , Ferguson, R. (1993). Auroral observations in the Antarctic at the time of the Tunguska event, 1908 June 30. // Australian journal of astronomy, Vol. 5, No. 1, pp. 1 – 10.

Stentzel, Arthur (1909). Die Dämmerungsanomalien im Sommer 1908. // Meteorologische Zeitschrift, v.26, pp. 437 - 446 (in German).

Strutt, R.J. (1931). On a night sky of exceptional brightness, and on the distinction between the polar aurora and the night sky. // Proceedings of the Royal Society of London. Series A, Vol. 131, No. 817, pp. 376-381 https://doi.org/10.1098/rspa.1931.0059

Süring, R. (1908). Die ungewöhnlichen Dämmerungserscheimlngen im

Juni und Juli 1908. // Bericht über die Tätigkeit des Preussischen Meteorologischen Instituts, pp.79-83 (in German).

Turco, R. P. , Toon, O. B. , Park, C. , Whitten, R. C. , Pollack, J. B. , Noerdlinger, P. (1982). An analysis of the physical, chemical, optical, and historical impacts of the 1908 Tunguska meteor fall. // Icarus, Volume 50, Issue 1, p. 1-52. https://doi.org/10.1016/0019-1035(82)90096-3

Vasil'ev, N.V., Zhuravlev, V.K., Zhuravleva, R.K., Kovalevskii, A.F., Plekhanov, G.F. (1965). Nochnye svetyaschiesya oblaka i opticheskie anomalii svyazannye s padeniem Tungusskogo meteorita. Nauka, Moskva, 112 p. (in Russian).

Vasilyev, N.V. (1998). The Tunguska Meteorite problem today. // Planetary and Space Science, Volume 46, Issues 2–3, pp. 129-150. https://doi.org/10.1016/S0032-0633(97)00145-1

Vasil'ev, N.V. (2004). Tungusskii meteorit. Kosmicheskii fenomen leta 1908 goda. M.: NP ID "Russkaya panorama", 372 p. (in Russian). ISBN 5-93165-106-3

Wolf, Max (1908). Über die Lichterscheinungen am Nachthimmel aus dem Anfang des Juli. // Astronomische Nachrichten,v.178, pp. 298 - 300 (in German).

Yakovlev, B.A. (1990). Atmosfernye osadki, grozy i Tungusskoe sobytie 1908 g. // Sledy kosmicheskikh vozdeistvii na Zemlyu. Novosibirsk, Nauka
, Sib. otd-nie, pp. 112 - 120 (in Russian).